\documentclass[preprint,review,11pt]{elsarticle}

\usepackage{graphicx}

\usepackage{amssymb}

\usepackage[top=4cm,left=3.9cm,right=3.9cm,bottom=3.8cm]{geometry}
\usepackage{csvsimple}
\usepackage{multirow}
\usepackage{lipsum}
\usepackage{afterpage}

\usepackage{etoolbox}
\robustify\bfseries

\usepackage{eurosym}
\usepackage{siunitx}

    \DeclareSIUnit\eur{\officialeuro}
    \DeclareSIUnit\M{M}
    \DeclareSIUnit\k{k}

  \def\sym#1{\ifmmode^{#1}\else\(^{#1}\)\fi}

\usepackage{setspace} 

\usepackage{csquotes}
\usepackage{amsmath}
\usepackage{bm}

\usepackage{xspace}
	\newcommand\ie{i.\,e.\xspace}
	\newcommand\eg{e.\,g.\xspace}

	\newcommand\US{US\xspace}
	\newcommand\EU{EU\xspace}

\usepackage[amsmath,hyperref,framed]{ntheorem} 
  \theoremstyle{plain}
     
  \theoremstyle{nonumberplain}
    \theoremseparator{.}
    \theoremheaderfont{\bfseries}
    \theorembodyfont{\normalfont}
    \theoremsymbol{$\blacksquare$}
      \RequirePackage{amssymb}

      \makeatletter
    \let\copy@theorem@headerfont=\theorem@headerfont
    \newcommand{\my@theorem@headerfont}{%
        \boldmath\copy@theorem@headerfont\unboldmath
      }
    \let\theorem@headerfont=\my@theorem@headerfont
      \makeatother

\theoremstyle{nonumberplain}
\theoremseparator{.}
\usepackage[%
  sort&compress
]{cleveref}
\usepackage{url}

\usepackage{isomath}

\usepackage{algorithmic}        

    \crefname{ALC@unique}{step}{steps}
    \Crefname{ALC@unique}{Step}{Steps}
    \crefname{ALC@line}{step}{steps}
    \Crefname{ALC@line}{Step}{Steps}

\usepackage[usenames,dvipsnames]{xcolor}

\usepackage{colortbl}
\usepackage{tabularx}
\usepackage{booktabs}
\usepackage{collcell}

\usepackage{ragged2e}
  
\newcommand{\PreserveBackslash}[1]{\let\temp=\\#1\let\\=\temp}
\newcolumntype{v}[1]{>{\PreserveBackslash\RaggedRight\hspace{0pt}}p{#1}}

  \usepackage{adjustbox}
\newcolumntype{Q}[2]{%
    >{\adjustbox{angle=#1,lap=\width-(#2)}\bgroup}%
    l%
    <{\egroup}%
}

\usepackage[
    colorinlistoftodos,
    textsize=footnotesize,
        ]{todonotes}

\makeatletter
    \renewcommand{\fps@figure}{htb}         
    \renewcommand{\fps@table}{htbp}         
\makeatother 

\usepackage{float}
\usepackage{rotating}

\usepackage{placeins}

\journal{Decision Support Systems}

\begin{document}

\begin{frontmatter}



\title{Long-term stock index forecasting based on text mining\\ of regulatory disclosures}


\author[ETH]{Stefan Feuerriegel\corref{cor1}}
\ead{sfeuerriegel@ethz.ch}

\author[Freiburg]{Julius Gordon}
\ead{juliusgordon89@gmail.com}

\address[ETH]{ETH Zurich, Weinbergstr. 56/58, 8092 Zurich, Switzerland}
\address[Freiburg]{University of Freiburg, Platz der Alten Synagoge, 79098 Freiburg, Germany}

\cortext[cor1]{Corresponding author.}

\begin{abstract}
Share valuations are known to adjust to new information entering the market, such as regulatory disclosures. We study whether the language of such news items can improve short-term and especially long-term (24 months) forecasts of stock indices. For this purpose, this work utilizes predictive models suited to high-dimensional data and specifically compares techniques for data-driven and knowledge-driven dimensionality reduction in order to avoid overfitting. Our experiments, based on 75,927 ad hoc announcements from 1996--2016, reveal the following results: in the long run, text-based models succeed in reducing forecast errors below baseline predictions from historic lags at a statistically significant level. Our research provides implications to business applications of decision-support in financial markets, especially given the growing prevalence of index ETFs (exchange traded funds). 
\end{abstract}


\end{frontmatter}



\section{Introduction}


The efficient market hypothesis formalizes how financial markets process and respond to new information \cite{Fama.1965}. Its semi-strong form states that asset prices fully reflect publicly-available information. Based on this premise, one can expect price changes whenever new information enters the market. In practice, regulations ensure that stock-relevant information is revealed primarily via regulatory disclosures in order to provide equal access for all market participants. Such materials disclose, for instance, quarterly earnings, but also management changes, legal risks and other events deemed important \cite{Li.2010b}. Accordingly, financial disclosures present an alluring and potentially financially-rewarding means of forecasting changes in stock valuations \cite{Nassirtoussi.2014}. 


In this respect, corporate news conveys a broad spectrum of information concerning the past performance and current challenges of the business \cite{Prollochs.2018b}, as well as frequently hinting at the future outlook. Research has followed this reasoning and empirically quantified the impact of the narrative content on the subsequent stock market responses \cite[cf.][]{Fisher.2016,Kearney.2014,Loughran.2016}. Moreover, researchers have also demonstrated the prognostic capability of financial disclosures with respect to individual stock market returns in the short term \cite[e.\,g.][]{Kraus.2017,
Wang.2012,Siering.2013b}. 
 Accordingly, news-based forecasting has received considerable traction and, as a result, various publications have evaluated different news datasets, forecasted indicator/markets, preprocessing operations from the field of natural language processing and forecasting algorithms. Here we refer to the literature, which provides a thorough overview \cite{Nassirtoussi.2014}. 

Forecasting the development of stock indices is highly demanded by multiple stakeholders in financial markets. The underlying reason is that households are investing their money no only in individual stocks, government bonds or savings accounts; rather, they increasingly prefer exchange-traded funds~(ETFs). These ETFs replicate the movements of marketable securities, with stock indices being the most prominent example. As part of their benefits, ETFs are traded on stock exchanges but often with higher liquidity and lower fees. Hence, private investors demand for decision support in better understanding the development of markets, as well as for obtaining prognostic support. For instance, more than 1700 different index ETFs have emerged, amouting to total assets worth more than USD~2.1~trillion.\footnote{ETF Daily News. \emph{ETF Stats For April 2015; Product Count Tops 1700}. URL: \url{https://etfdailynews.com/2015/05/18/etf-stats-for-april-2015-product-count-tops-1700/}, accessed April~29, 2018.}

While previous studies provide empirical evidence suggesting a link between financial disclosures and stock index dynamics in the short run, further research is needed to investigate the possibility of long-term forecasting. In this regard, a recent literature review reveals that evidence concerning the long-term prognostic power of financial news is scarce \cite{Nassirtoussi.2014}. As a remedy, it presents the object of this paper to investigate the predictive capacity of regulatory disclosures in forecasting future index levels in the long term. This undertaking seems especially relevant for practitioners in, for example, monetary policy and the investment industry as their decision-making is based on the economic outlook, as is reflected by market indices. 

Despite these aforementioned investigations, it has yet to be established whether financial disclosures can facilitate the long-term forecasting of stock indices. For this purpose, we need to make provisions for the high-dimensional predictor matrices that arise in text mining and thus experiment with different methods from machine learning that are carefully chosen for our setting. This presents a challenging undertaking that is often referred to as \textquote{wide} data, since the presence of single words entails only little prognostic power and, in addition, we face a higher number of predictors than observations, which must be effectively handled. This increases the risk of overfitting and we thus show that dimensionality reduction can provide effective means to overcome this problem. 

The novelty of this work is to apply text mining procedures in order to evaluate long-term forecasts of stock indices with wide predictor matrices. We specifically run experiments with (1) machine learning and high-dimensional news. We further extend these models by means of additional feature reduction in order to reduce the risk of overfitting through data-driven dimensionality reduction. (2)~We aggregate different sentiment scores and then insert these into our machine learning models. This represents a form of explicit feature engineering as part of a knowledge-driven dimensionality reduction. (3)~We perform an a priori reduction process in order to filter news from large-cap firms as these might be more relevant. Altogether, the extensive set of experiments yields prescriptive recommendations for implementing powerful news-based forecasts. 

In this work, we utilize 75,927 regulatory ad~hoc announcements in German and English together with three different stock indices, namely, the German prime index~(DAX), the German composite index~(CDAX), and the STOXX Europe~600. Our text-based models include extreme gradient boosting, principal component regression and the random forest, as well as the elastic net with it special cases, the lasso and ridge regression. Our models are compared to different techniques for linear and non-linear autoregression that serve as our baselines. We note that our implementation was carefully designed to circumvent a potential look-ahead bias \cite{Neuhierl.2011,White.2000}, which would incorporate variables that are not present at the time of the forecast. While not all predictive experiments outperform the baselines, our evaluations still reveal the promising performance of our text-based predictions, especially for the long-term forecasts. 


Our proposed text mining approach provides decision support for financial markets and thus entails a number of implications for management and individuals. On the one hand, our machine learning framework contributes to automated trading in financial markets. It also helps managers from institutional trading in making profitable investment decisions. On the other hand, it even facilitates retail investors, such as individual investors from online trading platforms, in managing their portfolio. Given the prevalence of ETFs as a widespread investment instrument for private households, the findings of this work have thus also direct implications to this group of stakeholders. 

The remainder of this paper is organized as follows. The above introduction has outlined a research gap concerning the long-term forecasts of stock indices based on the language embedded in corporate disclosures. To address this issue, we review related work (\Cref{sec:related_work}) and introduce our text mining models for news-based forecasting in \Cref{sec:methods} and our datasets in \Cref{sec:datasets}. \Cref{sec:results} then measures the improvements of our text-based forecasts over time series models with autoregressive terms. Based on the findings, \Cref{sec:implications} discusses the implications of our work, while \Cref{sec:conclusion} concludes.

\section{Related work} 
\label{sec:related_work}

\subsection{Decision support from financial news}

News-based predictions have become a common theme in decision support literature, yet rarely with a focus on stock indices. Hence, we decided to present a relatively broad overview that illustrates examples from the different streams in previous research. For a complete overview, we refer to the survey of predictive text mining for financial news in \cite{Nassirtoussi.2014,Ravi.2015}. Accordingly, related works evaluate different (1)~forecasting algorithms from the field of natural language, (2)~forecasted market variables and (3)~news datasets. These are outlined in the following.


The underlying algorithms are often named opinion mining or sentiment analysis, consistent with the terminology in computational natural language processing \cite{Pang.2008}. These can extract both the fundamental and qualitative information that forms the foundation for the decision-making of market stakeholders \cite{deFortuny.2014,Geva.2014,Hagenau.2013}. The underlying forecasting techniques usually follow the same procedure, where the first step pre-processes the running text and then transforms it into a mathematical representation that serves as input to a subsequent machine learning classifier \cite{Nassirtoussi.2014,Prollochs.2016b,Schumaker.2012}. Examples include support vector machines \cite{Fisher.2016,Hagenau.2013, Schumaker.2012}, decision tree classifiers \cite{Chan.2011}, artificial neural networks and boosting methods \cite{Chan.2011, Nassirtoussi.2014}. Alternatively, algorithms based on deep learning circumvent the manual need for feature engineering \cite{Kraus.2017}. Yet other works explicitly cater for the time-varying nature of sentiments \cite{Ho.2017}.
  

These approaches are used to forecast various indicators of interest. These include, for instance, nominal returns \cite{Hagenau.2013}, abnormal returns \cite{Kraus.2017}, optimal trading decisions \cite{Feuerriegel.2016}, and market volatility \cite{Groth.2011}. A recent contribution by \cite{Li.2014} applied an ontology-based web mining framework to improve the accuracy of unemployment rate predictions in the \US. In some cases, individual news are further enriched by the wisdom of crowds \cite{Eickhoff.2016}. Yet an even different stream of research is interested in macroeconomic indicators \cite{Feuerriegel.2018}.


Examples of news sources include newspaper articles from media sources, such as the Wall Street Journal, Bloomberg, Yahoo Finance \cite[e.\,g.][]{Schumaker.2012}; news wires; and regulated fillings, such as ad~hoc announcements, 8-K fillings and annual reports \cite[e.\,g.][]{Wang.2012, Feuerriegel.2016, Hagenau.2013, Prollochs.2016b, Prollochs.2018}. Additional sources cover alternative media, such as social media, user-generated content and microblogs \cite[e.\,g.][]{Deng.2017, Li.2013, Bollen.2011, Wuthrich.1998}. 

\subsection{Stock index forecasting}

The link between disclosure content and financial markets is found not only for individual stocks, but also in the case of stock indices. For instance, 
 the S\&500 index \cite{Zubair.2015} is positively correlated with specifically constructed sentiment metrics. In a predictive setting, media articles facilitate forecasting experiments that predict the same-day return of the Dow Jones index \cite{Siering.2013}. Similarly, the momentum of news tone seems capable of predicting the direction of CDAX movements \cite{Hagenau.2013c}. Here the predictions are made between one and ten weeks ahead, but this work lacks a rigorous comparison to baselines (\eg time series models) in order to convincingly demonstrate that the text-based forecast outperforms simple autoregressive models. Hence, the added value of news-based predictors remains unclear. 
 
In the context of this manuscript, a wide array of previous works have investigated the possibility of forecasting stock indices based merely on historic values and, hence, we point out only a few illustrative examples in the following. Forecasting experiments have been undertaken for various indices such as the S\&P~500 \cite{Leung.2000,Rounaghi.2016}, the NYSE \cite{Pai.2005}, the Dow Jones Industrial Average \cite{Wang.2012b}, the NIKKEI~225 \cite{Huang.2005} and also for emerging markets \cite{Oztekin.2016}. These works frequently utilize time series analysis methods, such as autoregressive or moving-average processes, occasionally together with approaches for volatility modeling \cite{Pai.2005,Rounaghi.2016,Herwartz.2017}. Works located in the proximity of machine learning also experiment, for instance, with support vector machines \cite{Huang.2005,Pai.2005}, neural networks \cite{Atsalakis.2009,Leung.2000,Oztekin.2016}, and hybrid models of neural networks and autoregression \cite{Wang.2012b}. Hence, we utilize both autoregressive and non-linear machine learning models with historic values as our benchmarks. 

Previous research \cite{Timmermann.2004} also provides evidence as to why our approach is likely to be superior, \ie the constant-parameters in classical time series analysis are designed to capture stationary processes, while our approach can even model the event-driven jump due to corporate disclosures.



%

\section{Text mining framework} 
\label{sec:methods}

This section details the forecasting models that serve as our baselines, as well as the models based on the content of financial disclosures. These aim at forecasting a financial time series $Y_t$ with $t = 1, \ldots, T$. The common challenge behind the following procedures is that the predictor matrix is extremely wide, which leads to the risk of overfitting. Hence, we overcome this problem by data-driven dimensionality reduction and explicit feature engineering with domain knowledge through sentiment analysis.


\subsection{Baselines with lagged data}

We implement a linear autogressive model~(lm) in order to forecast future observations from the historic time series, where the variable $l$ refers to the number of lags, $t$ to current time step and $t+h$ to the forecasted time step when making a prediction $h$ steps ahead. The autoregressive process is then modeled via  
\begin{equation} 
\label{eq:AR}
Y_{t+h} = {\alpha} + {\beta}_1 Y_{t-1} + \ldots + {\beta}_l Y_{t-l} + {\varepsilon}_i ,
\end{equation}
with coefficients $\alpha, \beta_1, \ldots, \beta_l$. It thus expects predictors $Y_{t-1}, \ldots, Y_{t-l}$ in order to forecast ${Y}_{t+h}$. 

We also apply machine learning models to the input vector $\left[ Y_{t-1}, \ldots, Y_{t-l} \right]^T$ with $l$ lags. This allows us to relax the assumption of a linear relationship and specifically test for non-linear dependencies. In this regard, we choose the same set of machine learning models as in the case of text-based approaches, namely, least squares absolute shrinkage operator~(lasso), ridge regression, elastic net~(enet), gradient boosting~(gbm), principal component regression~(pcr) and random forest~(rf).

\subsection{Sentiment-based machine learning}

Predictions from several hundred documents as a single observation have oftentimes demonstrated to increase the risk of overfitting. A viable trade-off is commonly presented by drawing upon sentiment dictionaries as a form of feature engineering. Here the idea is to incorporate domain knowledge in the form of pre-defined dictionaries that label terms into different semantic categories \citep{Prollochs.2015c}. The conventional assumption is that the overall sentiment hints the economic outlook \citep{Tetlock.2007,Tetlock.2008,Loughran.2016}. Sentiment dictionaries have frequently been utilized in explanatory research where the objective is to identify a statistically significant relationship between content of financial disclosures and the corresponding stock market reaction \citep[e.\,g.][]{Ho.2017,Hagenau.2013c}, yet empirical evidence on their potential advantage in long-term predictive settings is scarce.

We create a predictor matrix consisting out of $l$ autoregressive lags and additional scores. Here we experiment with three approaches:
\begin{enumerate}
\item We compute a single sentiment score that reflects the overall polarity of the language. This is given by the relative ratio between the number of positive and negative words, \ie $\cfrac{\#\text{positive} - \#\text{negative}}{\#total}$. 
\item Beyond that, we also compute separate scores measuring the use of positive and negative language. These are formalized by ratios $\cfrac{\#\text{positive}}{\#total}$ and $\cfrac{\#\text{negative}}{\#total}$.
\item Sentiment dictionaries often consist of further categories, such as uncertainty expressions (especially with regard to the economic climate). Hence, we extend our previous positivity and negativity metrics by a proportional score of uncertainty words.
\end{enumerate}

The computational advantage of these approaches is that even long narrative materials can be easily mapped on a numerical figure that can adapt to the underlying valence of the tone, as well as the economic outlook. At the same time, this procedure reduces the degrees-of-freedom immensely and instead, replaces these by domain knowledge as encoded in the dictionaries, thereby diminishing the potential of overfitting.

In all of our experiments, we incorporate the Loughran-McDonald finance-specific dictionary\footnote{URL: \url{https://www3.nd.edu/~mcdonald/Word_Lists.html}, last accessed April 3, 2018.} which has evolved as a quasi-standard in finance-related research \citep{Loughran.2016}. This dictionary has specifically constructed such that they can extract qualitative materials from financial news in order to yield numerical scores.

\subsection{Text-based machine learning}


Text-based require that the running text is transformed in a machine-ready representation, to which one can later apply a machine learning classifier \cite{Manning.1999}. For this reason, the conventional bag-of-words approach is to process the original document by counting the frequency of tokens \cite{Nassirtoussi.2014,Pang.2008,Ravi.2015}. These frequencies can optionally be weighted, until they finally serve as features in the machine learning models. 
 

\begin{figure}[H]
\centering
\includegraphics[width=\linewidth]{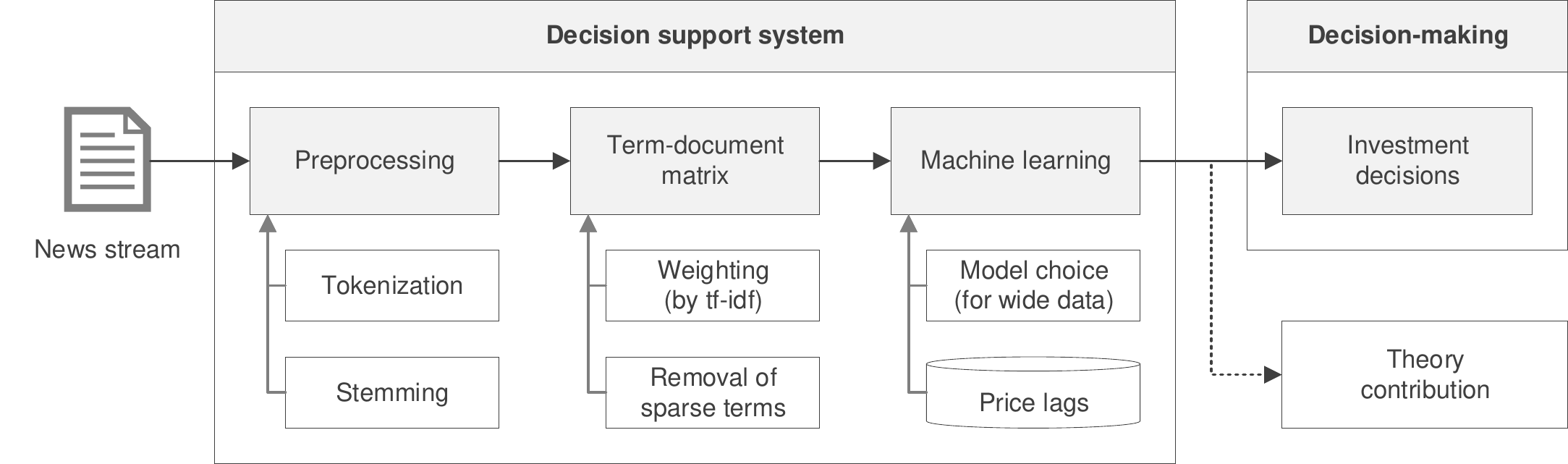}
\caption{Text mining framework that serves as the foundation of a decision support system for facilitating decision-making in financial markets.}
\label{fig:flow_diagram}
\end{figure}

Our text-based models adhere to the previous approach and we thus detail our methodology in the following (cf. \Cref{fig:flow_diagram}). We remove numbers, punctuations, and stop words\footnote{Here we follow the suggested list of the package \textquote{tm} in R.}, followed by stemming. We then count the frequency of all terms appearing in the disclosures belonging to each time step $t = 1, \ldots, T$. This results in a document-term matrix $X \in\mathbb{R}^{T\times P}$ where the columns refer to the different frequencies of all $P$ terms. Hence, one row denotes the frequency of the terms in the disclosures of a single time step $t$. We subsequently scale the matrix by the relative informativeness of words as defined by the tf-idf weighting \cite{Manning.1999}. The resulting rows then serves as the main predictors for future values $Y_{t+h}$. 

The corpus is further processed in order to yield high-dimensional predictor matrices as detailed in the following. More precisely, the document-term matrix $X$ entails an extremely wide format, wherein the number of predictors exceeds the observations by far. We thus follow a heuristic approach for reducing the dimensionality of $X$ further. That is, we omit rare terms for which the corresponding columns contain more than 10\,\% of sparse entries to reduce the risk of overfitting, as well as the necessary computational resources.

Care is necessary when choosing a suitable machine learning model, since we require model that can generalize well even with more predictors than data samples. In other words, we face a situation where the number of words exceeds the number of past observations, which can easily result in overfitting for many machine learning models. As a remedy, we decide upon predictive models that are perform known to handle such wide datasets effectively \cite{Hastie.2013}: lasso, ridge regression, elastic net, gradient boosting, principal component regression and random forest, which uilize implicit feature selection, regularization or dimensionality reduction in order to yield a favorable bias-variance tradeoff and thus avoid potential overfitting. 

Beyond tf-idf features, we also experiment with alternative approaches as part of our feature engineering in order to reduce the risk of overfitting. That is, we apply techniques for unsupervised dimensionality reduction: the document-term matrix $X$ is replaced by (a)~its principal component analysis~(pca) and (b)~a latent semantic analysis~(lsa).

Subsequently, we evaluate two distinct strategies to news-based forecasts: (1)~the above document-term matrices or its transformations serve as the sole predictor (\ie $l = 0$). (2)~The document term-matrices are further augmented by $l$ autoregressive terms from the predicted stock index with $l = 1$ or $l = 6$. As a result, the latter approach further adapt to seasonalities and short-term trends. Altogether, this yields 63 different predictive models as subject for our experiments (\ie 7 classifiers, with raw tf-idf and 2 adaptations, namely, pca and lsa; each with three choices of $l$).

\subsection{Parameter calibration}

We proceed as follows in order to tune the hyperparameters of our predictive models. For that purpose, we chronologically split the dataset into two subsets for training (60\,\% of observations) and testing (remaining 40\,\%) in order to preserve the temporal order of the disclosures. We then find the best-performing parameters by performing a grid search. More specifically, we utilize time-series cross-validation with a rolling forecast origin \cite{Hyndman.2014}. This procedure first splits the training data $\mathcal{T}$ into $k$ disjoint subsets $\mathcal{T}_1, \ldots, \mathcal{T}_k$ in temporal order. We then iterate over all possible combinations of the tuning ranges for each parameter and all values of $i = 2, \ldots, k$. For each combination, we learn the model parameters from the previous subsets in time given by $\mathcal{T}_1, \ldots, \mathcal{T}_{i-1}$. Subsequently, we compute the performance of this model calibration on the validation set $\mathcal{T}_i$. Finally, we return the best-performing hyperparameter setting. 

Throughout this paper, all computational experiments are performed by utilizing $k = 10$ splits. We rely upon the default search grid as defined by the \textquote{caret} package in R for reasons of comparability \cite{Kuhn.2008}.

\section{Datasets}
\label{sec:datasets}

\subsection{Regulatory disclosures}

Our dataset of regulatory disclosures contains all ad~hoc announcements that were disseminated by the DGAP (Deutsche Gesellschaft fuer Ad-hoc-Publizitaet), a subsidy of EQS Group, which is the leading publishing service provider for mandatory ad~hoc announcements in Germany. Government regulations require firms to publish all stock-relevant materials first and without delay via this channel. The disclosure of these filings is obligated by regulatory policies according to the German Securities Trade Act and affects all firms listed on German stock exchanges. As a result, the rules not only apply for German firms but also for foreign ones listed there, which is the reason why these disclosures are typically published in German, English or both. We thus specifically compare the prognostic capabilities of the aforementioned languages in predicting stock prices. The choice of this dataset entails a number of practical advantages. First, the strict publication rules warrant a timely publication and ensure that the content is relevant to stock markets. This prohibits firms from disseminating the information via the press before filing an ad~hoc disclosure. Second, each ad~hoc announcement must be signed by the head of the company. Third, the quality of all filings is further quality-checked by the Federal Financial Supervisory Authority (\ie BaFin).

We collected all 80,813 ad~hoc announcements from July~1996 through April~2016 that were disseminated by the DGAP.\footnote{The dataset is available upon request via email.} These materials were retrieved via the dedicated online channel (\url{http://www.dgap.de/dgap/News/?newsType=ADHOC}). We specifically note that the dataset underwent no additional (subjective) filtering steps in order eliminate the risk of data dredging and associated look-ahead biases \cite{Neuhierl.2011,White.2000}, which would incorporate information that are not present at the time of the forecast.

Empirical evidence has demonstrated a strong response of share prices in the wake of this type of financial news, as well as a high prognostic capability of changes in stock valuations \cite{
Hagenau.2013c
}. Moreover, the content of such disclosures also enables previous works in predicting volatility and risk-related metrics \cite{Groth.2014
}. Altogether, this indicates that ad~hoc announcements are likely to give an accurate sense of current developments for individual firms, in addition to reflecting the market environment. 





\subsection{Stock index data}

Since ad~hoc announcements can originate from German or even foreign corporations, we have to reflect this fact and make a corresponding choice of stock indices (see \Cref{tbl:variables}). We thus incorporate two German stock indices: the DAX includes the \num{30} biggest stocks trading on the Frankfurt stock exchange, while the CDAX consists of all German stocks listed in the general standard or prime standard market segments, which totals to approximately 485 firms. In addition, we experiment with the STOXX Europe~600 as another point of comparison. We collected these financial time series from Bloomberg in both weekly and monthly resolutions. This amounts to 1,087 weekly and \num{260} monthly observations.

\begin{table}[H] 
\centering 
\footnotesize
\onehalfspacing
\makebox[\textwidth]{
\begin{tabular}{l p{5cm} l p{2.3cm} p{4.5cm}} 
\toprule
\textbf{Symbol} & \textbf{Name} & \textbf{Region} & \textbf{Resolutions (in use)} & \textbf{Notes} \\ 
\midrule
\textbf{DAX} & German prime stock index & Germany & Monthly \&~weekly & Index of \num{30} selected German blue chip stocks\\
\textbf{CDAX} & German composite stock index & Germany & Monthly \&~weekly & Composite index of all stocks traded on the Frankfurt Stock\\
\textbf{STOXX} & STOXX Europe 600 & \EU & Monthly \&~weekly & Composite index from across the European region \\
\bottomrule
\end{tabular}%
}%
\caption{Overview of predicted stock indices.} 
\label{tbl:variables} 
\end{table}

\subsection{Summary statistics}

\begin{table}[H] 
\centering 
\footnotesize
\singlespacing
\begin{tabular}{l SS SSS} 
\toprule
\textbf{Year} & \textbf{Disclosures} & \textbf{Mean length} & \textbf{DAX} & \textbf{CDAX} & \textbf{STOXX} \\ 
\midrule
1996 (Jul--Dec) & 424 & 134.08 & 2591.79 & 240.79 & 154.80 \\ 
1997 & 1523 & 136.55 & 3744.49 & 335.59 & 211.79 \\ 
1998 & 1911 & 167.12 & 5058.75 & 435.16 & 272.84 \\ 
1999 & 3863 & 218.98 & 5391.62 & 457.76 & 312.62 \\ 
2000 & 6954 & 247.26 & 7049.20 & 578.83 & 379.73 \\ 
2001 & 8814 & 184.30 & 5612.18 & 448.99 & 314.87 \\ 
2002 & 4983 & 178.78 & 4111.16 & 345.18 & 249.62 \\ 
2003 & 4676 & 189.64 & 3205.03 & 280.41 & 204.03 \\ 
2004 & 4095 & 193.73 & 3984.07 & 351.00 & 239.83 \\ 
2005 & 4112 & 204.08 & 4706.41 & 418.40 & 279.14 \\ 
2006 & 4221 & 218.66 & 5962.27 & 536.32 & 336.00 \\ 
2007 & 4449 & 237.32 & 7563.47 & 682.81 & 378.89 \\ 
2008 & 4012 & 241.40 & 6149.94 & 546.43 & 278.03 \\ 
2009 & 3500 & 253.12 & 5021.33 & 433.34 & 215.69 \\ 
2010 & 3054 & 274.83 & 6161.05 & 538.48 & 256.21 \\ 
2011 & 2982 & 296.49 & 6679.14 & 589.70 & 261.42 \\ 
2012 & 2903 & 304.52 & 6911.78 & 611.16 & 262.67 \\ 
2013 & 3099 & 289.79 & 8374.98 & 748.53 & 303.06 \\ 
2014 & 3031 & 295.35 & 9616.60 & 860.55 & 338.88 \\ 
2015 & 2811 & 306.32 & 11006.63 & 993.91 & 380.60 \\ 
2016 (Jan--Apr) & 510 & 314.09 & 10021.38 & 921.06 & 339.75 \\ 
\bottomrule
\end{tabular}%
\caption{Summary statistics of corporate disclosures, as well as the annual mean of the stock index data.} 
\label{tbl:descriptives} 
\end{table}

\Cref{tbl:descriptives} provides summary statistics related to our dataset. On average, each ad~hoc announcement contains \num{232.7} words, while we see a slight upward trend across time. Our corpus thus entails a total of \num{17.7} million terms. The annual mean number of disclosures is 3,947 for the time frame covering 1997--2015. Finally, we note a high correlation coefficient between DAX and CDAX close to \num{1}. The correlation with the STOXX Europe~600 remains below that value, amounting to \num{0.80}~for the DAX and \num{0.77}~in the case of the CDAX.


%

\section{Results} 
\label{sec:results}


This section describes the setup of our computational experiments, for which it then reports the results of the out-of-sample forecasting.

\subsection{Computational setup}

The purpose of our experiments is to compare the predictive performance of the benchmark models to the disclosure-based forecasts. We thus train models with the raw time series of each stock index. Here, we run our experiments by setting the number of lags to $l = 6$ in order to provide a reasonable trade-off between bias and variance. This choice yields fairly stationary subsets and has also been utilized by previous works \cite[e.\,g.][]{Litterman.1986,Stock.2002}. We also study the sensitivity by performing experiments with a single lag ($l = 1$) as a comparison. Finally, we incorporate our text-based, high-dimensional predictor matrix and train them both without lags and, consistent with above, with $l = 6$ lags. 

Across all our experiments, we specifically forecast the raw values given by $Y_{t+h}$ without further transformations. We explicitly refrain from using transformations, as practitioners are interested in the actual values and this thus presents a more realistic setting. The variable $Y_{t+h}$ is easily interpretable and especially demanded by practitioners. We then make predictions across different forecast horizons $h$. Here we draw upon different horizons $h$ in case of monthly and weekly resolution. All of the aforementioned values refer to a maximum forecast horizon of 24~months in both cases. The prediction for $h = 1$ is modeled as first-differences as this appears to better identify turning points in the business cycle. 

In the following, we quantify the forecast performance based on the root mean squared error~(RMSE). \Cref{tbl:results_monthly} reports the results for the monthly time series and \Cref{tbl:results_weekly} for the weekly one. Furthermore, we follow the recommendations in \cite{Giacomini.2013} by running a Diebold-Mariano~(DM) test in order to ensure the robustness of our findings. Given a certain sample, the null hypothesis tests whether forecasts with the text-based predictor matrix are at least as accurate as forecasts lacking these external inputs \cite{Diebold.1995,Diebold.2015}. The corresponding statistic thus reveals whether the reduction of forecast errors is statistically significant. Here we take the squared-error as the loss function. Accordingly, this provides statistical confidence regarding the advantages of utilizing the text-based models over the benchmarks with merely lagged input values for the given test data.

Detailed results by model are listed in the supplements. These contribute to the robustness of the proposed machine learning approach. Oftentimes, these predictions yield the same pattern as in the summarizing table, since, when the baseline is clearly surpassed there, we can outperform the best benchmark in both the nowcasting and the long-term scenario for the majority of machine learning models.

We further conduct the following sensitivity check. That is, we assume that the contribution of firms to the overall market movements is linked to their market capitalization. Hence, we expect the stock indices to be particularly moved by large-cap companies and thus filter our for the top-25 companies by market capitalization.

\begin{table}[!htbp]
\centering
\makebox[\textwidth]{
\tiny
\sisetup{input-symbols=(),round-mode=places,round-precision=3}
\begin{tabular}{@{\extracolsep{5pt}} l SSS} 
\toprule
\textbf{Input/model} & \multicolumn{1}{c}{$\bm{h=1}$} & \multicolumn{1}{c}{$\bm{h=12}$} & \multicolumn{1}{c}{$\bm{h=24}$}\\ 
\midrule
\multicolumn{4}{c}{Predicted variable: monthly German prime index (DAX)} \\
\midrule
  Benchmark: lags & 429.656 & 2232.505 & 3700.613 \\ 
   & {lm1} & {lm1} & {lm1} \\[0.3em] 
  Sentiment & \bfseries 421.925 & 2241.000 & \bfseries 3453.951 \\ 
   & \bfseries (0.047) & (1.000) & \bfseries (0.000) \\ 
   & \bfseries {Pos\&neg} & {Pos\&neg} & \bfseries {Sentiment} \\[0.3em] 
  Machine learning & \bfseries 409.662 & 2486.562 & \bfseries 3089.563 \\ 
   & \bfseries (0.006) & (0.998) & \bfseries (0.000) \\ 
   & \bfseries {ridge1} & {pca-gbm1} & \bfseries {rf6} \\[0.3em] 
	Incl. dimensionality reduction & \bfseries 411.062 & 2486.562 & \bfseries 3356.676 \\ 
   & \bfseries (0.034) & (0.998) & \bfseries (0.000) \\ 
   & \bfseries {pca-glmnet} & {pca-gbm1} & \bfseries {pca-gbm1} \\[0.3em] 
  Sensitivity: top-25 firms & \bfseries 409.662 & 2507.928 & \bfseries 3026.000 \\ 
   & \bfseries (0.006) & (1.000) & \bfseries (0.000) \\ 
   & \bfseries {ridge1} & {pca-gbm1} & \bfseries {rf1} \\[0.3em] 
  Sensitivity: corpus & \bfseries 410.475 & 2640.812 & \bfseries 2977.528 \\
   & \bfseries (0.115) & (1.000) & \bfseries (0.000) \\ 
   & \bfseries {Complete} & {Complete} & \bfseries {German} \\ 
\midrule
\multicolumn{4}{c}{Predicted variable: monthly German composite index (CDAX)} \\
\midrule
  Benchmark: lags & 36.919 & 204.343 & 339.495 \\ 
   & {gbm1} & {lm1} & {lm1} \\[0.3em] 
  Sentiment & \bfseries 36.486 & 205.118 & \bfseries 316.701 \\ 
   & \bfseries (0.223) & (1.000) & \bfseries (0.000) \\ 
   & \bfseries {Pos\&neg} & {Pos\&neg} & \bfseries {Pos\&neg} \\[0.3em] 
  Machine learning & \bfseries 35.376 & 236.667 & \bfseries 283.987 \\ 
   & \bfseries (0.003) & (1.000) & \bfseries (0.000) \\ 
   & \bfseries {ridge1} & {ridge6} & \bfseries {rf1} \\[0.3em] 
	Incl. dimensionality reduction & \bfseries 35.273 & 245.063 & \bfseries 307.168 \\ 
   & \bfseries (0.017) & (1.000) & \bfseries (0.000) \\ 
   & \bfseries {pca-glmnet} & {pca-gbm1} & \bfseries {pca-gbm1} \\[0.3em] 
  Sensitivity: top-25 firms & \bfseries 35.273 & 229.416 & \bfseries 284.336 \\ 
   & \bfseries (0.003) & (1.000) & \bfseries (0.000) \\ 
   & \bfseries {pca-glmnet} & {ridge1} & \bfseries {rf} \\[0.3em] 
  Sensitivity: corpus & \bfseries 35.34 & 246.833 & \bfseries 273.436 \\ 
   & \bfseries (0.007) & (1.000) & \bfseries (0.000) \\ 
   & \bfseries {Complete} & {Complete} & \bfseries {Complete} \\ 
\midrule
\multicolumn{4}{c}{Predicted variable: monthly STOXX Europe 600 index} \\
\midrule
  Benchmark: lags & 12.539 & 36.696 & 56.26 \\ 
   & {ridge6} & {gbm1} & {lm1} \\[0.3em] 
  Sentiment & 12.551 & 39.954 & \bfseries 50.931 \\ 
   & (0.519) & (0.972) & \bfseries (0.001) \\ 
   & {Sentiment} & {Pos\&neg} & \bfseries {Sentiment} \\[0.3em] 
  Machine learning & \bfseries 12.17 & 43.967 & \bfseries 50.557 \\ 
   & \bfseries (0.120) & (0.989) & \bfseries (0.000) \\ 
   & \bfseries {ridge6} & {gbm} & \bfseries {gbm6} \\[0.3em] 
	Incl. dimensionality reduction & \bfseries 12.487 & 45.602 & \bfseries 50.654 \\ 
   & \bfseries (0.405) & (1.000) & \bfseries (0.000) \\ 
   & \bfseries {pca-rf6} & {pca-rf1} & \bfseries {lsa-pcr1} \\[0.3em]
  Sensitivity: top-25 firms & \bfseries 12.17 & 44.799 & \bfseries 50.654 \\ 
   & \bfseries (0.120) & (0.999) & \bfseries (0.000) \\ 
   & \bfseries {ridge6} & {pca-rf1} & \bfseries {lsa-pcr1} \\[0.3em] 
  Sensitivity: corpus & \bfseries 12.36 & 45.394 & \bfseries 36.226 \\ 
   & \bfseries (0.358) & (0.996) & \bfseries (0.000) \\ 
   & \bfseries {Complete} & {German} & \bfseries {German} \\ 
\bottomrule
\end{tabular}
}
\caption{Comparison of prediction performance (root mean squared error) across different monthly stock indices, where we make predictions $h$ time steps ahead. Only the best-in-breed model is listed, for which we add bold highlighting when the model is equal or superior to the baseline. The corresponding $P$-value from the Diebold-Mariano test is given in brackets, as well as the type of the final model.}
\label{tbl:results_monthly}  
\end{table}

\begin{table}[!htbp]
\centering
\makebox[\textwidth]{
\tiny
\sisetup{input-symbols=(),round-mode=places,round-precision=3}
\begin{tabular}{@{\extracolsep{5pt}} l SSS} 
\toprule
\textbf{Input/model} & \multicolumn{1}{c}{$\bm{h=1}$} & \multicolumn{1}{c}{$\bm{h=52}$} & \multicolumn{1}{c}{$\bm{h=104}$}\\ 
\midrule
\multicolumn{4}{c}{Predicted variable: weekly German prime index (DAX)} \\
\midrule
Benchmark: lags & 236.176 & 2250.741 & 3711.472 \\ 
& {glmnet6} & {lm1} & {lm1} \\[0.3em] 
Sentiment & 236.342 & 2329.701 & \bfseries 3555.981 \\ 
& (0.573) & (1.000) & \bfseries (0.000) \\ 
& {Pos\&neg} & {Pos\&neg} & \bfseries {Pos\&neg} \\[0.3em] 
Machine learning & 236.357 & 2510.681 & \bfseries 3457.280 \\ 
& (0.586) & (1.000) & \bfseries (0.000) \\ 
& {pcr6} & {gbm} & \bfseries {gbm6} \\[0.3em] 
Incl. dimensionality reduction & 236.188 & 2600.012 & \bfseries 3277.951 \\ 
& (0.503) & (1.000) & \bfseries (0.000) \\ 
& {pca-rf} & {pca-gbm1} & \bfseries {pca-ridge6} \\ 
Sensitivity: top-25 firms & \bfseries 235.816 & 2469.524 & \bfseries 3277.951 \\ 
& \bfseries (0.408) & (1.000) & \bfseries (0.000) \\ 
& \bfseries {pca-rf1} & {gbm} & \bfseries {pca-ridge6} \\[0.3em] 
Sensitivity: corpus & \bfseries 235.604 & 2396.174 & \bfseries 3188.030 \\ 
& \bfseries (0.309) & (1.000) & \bfseries (0.000) \\ 
& \bfseries {German} & {Complete} & \bfseries {German} \\ 
\midrule
\multicolumn{4}{c}{Predicted variable: weekly German composite index (CDAX)} \\
\midrule
Benchmark: lags & 20.257 & 206.621 & 341.331 \\ 
& {glmnet6} & {lm1} & {lm1} \\[0.3em] 
Sentiment & \bfseries 20.188 & 211.286 & \bfseries 333.229 \\ 
& \bfseries (0.285) & (1.000) & \bfseries (0.000) \\ 
& \bfseries {Sentiment} & {Pos\&neg} & \bfseries {Pos\&neg} \\[0.3em] 
Machine learning & 20.274 & 235.412 & \bfseries 328.827 \\ 
& (0.565) & (1.000) & \bfseries (0.000) \\ 
& {rf1} & {gbm} & \bfseries {lasso} \\[0.3em] 
Incl. dimensionality reduction & \bfseries 20.251 & 242.067 & \bfseries 310.824 \\ 
& \bfseries (0.486) & (1.000) & \bfseries (0.000) \\
& \bfseries {pca-gbm1} & {pca-ridge1} & \bfseries {pca-ridge6} \\[0.3em] 
Sensitivity: top-25 firms & 20.264 & 233.816 & \bfseries 310.824 \\ 
& (0.522) & (1.000) & \bfseries (0.000) \\ 
& {pca-rf} & {ridge6} & \bfseries {pca-ridge6} \\[0.3em] 
Sensitivity: corpus & \bfseries 20.18 & 220.32 & \bfseries 304.383 \\ 
& \bfseries (0.203) & (1.000) & \bfseries (0.000) \\ 
& \bfseries {German} & {Complete} & \bfseries {German} \\ 
\midrule
\multicolumn{4}{c}{Predicted variable: weekly STOXX Europe 600 index} \\
\midrule
Benchmark: lags & 7.854 & 39.371 & 58.989 \\ 
& {lm6} & {gbm1} & {lasso6} \\[0.3em] 
Sentiment & 7.882 & 41.618 & \bfseries 54.862 \\ 
& (0.781) & (0.989) & \bfseries (0.063) \\ 
& {All categories} & {Pos\&neg} & \bfseries {Sentiment} \\[0.3em] 
Machine learning & 7.891 & 43.888 & \bfseries 53.287 \\ 
& (0.844) & (0.992) & \bfseries (0.000) \\ 
& {pcr6} & {rf6} & \bfseries {pcr} \\[0.3em] 
Incl. dimensionality reduction & 7.891 & 46.934 & \bfseries 49.996 \\ 
& (0.844) & (1.000) & \bfseries (0.000) \\ 
& {pca-pcr6} & {pca-rf} & \bfseries {pca-rf6} \\[0.3em]
Sensitivity: top-25 firms & 7.891 & 43.366 & \bfseries 49.996 \\ 
& (0.775) & (0.984) & \bfseries (0.000) \\ 
& {pcr6} & {rf6} & \bfseries {pca-rf6} \\[0.3em] 
Sensitivity: corpus & 7.872 & 44.133 & \bfseries 48.318 \\ 
& (0.632) & (0.998) & \bfseries (0.000) \\ 
& {Complete} & {Complete} & \bfseries {Complete} \\ 
\bottomrule
\end{tabular}
}
\caption{Comparison of prediction performance (root mean squared error) across different weekly stock indices, where we make predictions $h$ time steps ahead. Only the best-in-breed model is listed, for which we add bold highlighting when the model is equal or superior to the baseline. The corresponding $P$-value from the Diebold-Mariano test is given in brackets, as well as the type of the final model.}
\label{tbl:results_weekly}  
\end{table}

\subsection{Prognostic power of textual materials}

We now provide statistics concerning the prognostic power of news content. On the one hand, this establishes the overall relevance of textual cues as potential predictors and, on the other hand, summarizes the difficulties of the research setup: several thousands of different words can theoretically provide hindsight of future price changes, yet only a fairly small set of observations are available. This directly leads to the risk of overfitting and reveals the inherent methodological challenge, since the wide predictor matrix requires appropriate dimensionality reduction. 

In the following, we draw upon the information-fusion-based sensitivity analysis \cite{Oztekin.2013}, which has been widely used in the decision support literature as a tool for quantifying the relevance of predictors \cite{Dag.2017,Delen.2012}. It essentially measures the change in RMSE when omitting or including a single variable in an ensemble of all models. We computed the information-fusion-based sensitivity analysis for the text-based predictor matrix in the setting with a one-step ahead prediction of the DAX as the outcome variable. Here we yield an average sensitivity score of \num{1.001} for all textual cues with a standard deviation of \num{0.017}. As a comparison, the lags attain sensitivity scores of up to \num{1.451}. This demonstrates that only few variables have strong prognostic capacity of the outcome variable and it is thus a challenge to identify this subset. As a remedy, this paper compares different strategies of dimensionality reduction that either follow a data-driven logic or additionally incorporate domain knowledge.

Based on the above discussion, we later expect that, in some cases, the text-based prediction models can even be inferior to the simple baseslines. This can happen when the dimensionality reduction has not been able to identify the subset of relevant predictors and, instead, has overfitted.

		%
%

\subsection{German prime index: DAX}

For the monthly data, the disclosure-based models surpass the forecast accuracy of the benchmark models for the \num{1} and \num{24}-months-ahead prediction. For the short-term horizon, the models from machine learning and large-cap firms prove to be the most accurate, achieving an RMSE of \num{409.662}. In comparison, the best benchmark model recorded an RMSE of \num{429.656}. For the long-term prediction horizon, the disclosure-based models prove again to be superior. The clear standout is given by the combined corpus with an RMSE of \num{2977.528}. This yields a significant improvement over the best performing benchmark with an RMSE of \num{3700.613}.

We find a similar pattern for the weekly resolution. A majority of the disclosure-based models are able to outperform the benchmark models. The combined corpus achieved the lowest RMSE for the one-step-ahead prediction of \num{235.604}. Further, improvements are also attained for the 2-year-ahead horizon. The corpus sensitivity model again proves to be the most accurate with an RMSE of \num{3188.030}. However, the benchmark model returns superior forecasts for the medium-term prediction horizon of \num{52} weeks. 

\subsection{German composite index: CDAX}

The results of the predictive experiments undertaken for the CDAX index are as follows. For the monthly prediction experiments, the disclosure-based models outperform the benchmark over both short and long-term prediction horizons. The reduction to large-caps and the data-driven dimensionality reduction proved to be the most accurate with a RMSE of \num{35.273}. In comparison, the best performing benchmark recorded a RMSE of \num{36.919}. For the long-term prediction horizon the results indicate that the disclosure-based models are proven again to be superior. The best result is obtained by when utilizing the combined corpus with an RMSE of \num{273.436} for the \num{24}-months-ahead horizon. This a significant improvement over RMSE of \num{339.495} recorded by the best performing benchmark over the same period. The results for the \num{12}-months-ahead horizon indicate a similar result to the DAX index that the disclosure-based models were unable to out-predict the benchmark.

The RMSE values for the weekly resolution of the CDAX index illustrate that a majority of the disclosure-based models are able to outperform the benchmark over a number of prediction horizons. The corpus sensitivity again attained the best RMSE of \num{20.180} for the \num{1}-week-ahead prediction horizon. The sentiment and dimensionality reduction also outperformed the benchmark RMSE of \num{20.257}. For the one-year-ahead horizon, no disclosure-based model was able to outperform the benchmark. Finally, for the 2-year-ahead horizon, all machine learning approaches were able to record lower forecast errors than the benchmark models. The RMSE of \num{304.383} achieved by the German corpus significantly outperforms the best benchmark model RMSE of \num{341.331}.

\subsection{STOXX Europe 600 index}

We now discuss the RMSE values from the prediction experiments undertaken for the monthly resolution of the STOXX \num{600} index. A majority of the disclosure-based models surpassed the forecast accuracy of the benchmark for the one-step-ahead horizon. The models with machine learning and large-cap filtering proved to be the most accurate with both models recording a RMSE of \num{12.170}. In comparison, the benchmark RMSE was \num{12.539}. Further improvements in predictive accuracy over the benchmark were achieved for the 2-year-ahead horizon. This time the use of only German news obtained the most accurate prediction with a RMSE of \num{36.226}; however, the best-of-breed results from all other approaches also recorded a better RMSE than that of the best benchmark model (\num{36.226}). In a similar result to the previous mentioned experiments for the DAX and CDAX, the benchmark models were more accurate in the medium run.

The results from the weekly prediction experiments show small variations to the previous patterns. The disclosure-based models are unable to outperform the benchmark over the short-term prediction horizons. However, we find evidence of predictability in the long run. Here we noted a lowest forecast error for the combined corpus. This is in line with our expectations as the STOXX index contains firms from all over Europe that thus might prefer reporting not only in German but also in English. 

\subsection{Comparison}

The overall performance of disclosure-based forecasts varies depending on the predicted variable, forecast horizon, model choice and input choice. While the performance of disclosure-based forecasts is not superior across all experiments, we point to the following cases wherein disclosures, as a matter of fact, yield significant reductions in forecast errors. In this regard, we identify our primary finding: the text-based models help in improving the long-term forecasts of the three stock indices. Here the explicit sentiment, implicit dimensionality reduction, machine learning and sensitivity models are able over various forecast horizons to strongly outperform the benchmark. In the case of monthly data, we obtain reductions in the RMSE by \SI{4.6}{\percent} for the DAX, \SI{4.4}{\percent} for the CDAX, and \SI{2.9}{\percent} for the STOXX Europe 600 for the short-term horizon. For the long-term horizon, we see reductions in RMSE of \SI{19.5}{\percent} for the DAX, \SI{19.4}{\percent} for the CDAX, and \SI{35.6}{\percent} for the STOXX Europe 600. 

The improvements step from different model choices. In the long run, the best results are often achieved by combined corpus as firms can utilize different languages for their reporting. Conversely, the short-term predictions largely benefit from machine learning, optionally a restriction on large-caps as part of reducing the complexity of the input. These even appear beneficial over data-driven techniques for dimensionality reduction, such as the principal component analysis. Interestingly, explicit dimensionality reduction facilitates the weekly resolution, while it impedes the monthly resolution. Sentiment-based approaches can outperform lag-based predictions, yet are themselves outperformed by other text-based machine learning. We further observe that oftentimes simple linear relationships as in the lasso appear among the best-in-breed model. A potential reason is that this type of model benefits from additional implicit reduction of the feature space and the parameters are fairly easy to calibrate, thus yielding a more robust model. 

We further compare the normalized RMSE in \Cref{tbl:nrmse}, which allows comparison across different scales in the outcome variable. We find that the STOXX Europe index has lower relative prediction errors than the German indices. A potential reason could stem from the fact that the ad~hoc announcements cover not only domestic corporations but also foreign firms listed at German stock exchanges. We also see a higher prognostic capacity for the DAX as compared to the composite index. A possible explanation could be that it is fairly difficult to accurately assess the wealth of information concerning all small-cap firms, thus leaving an additional component of stock dynamics that entails considerable variability and thus cannot be fully explained with our current models. 

\begin{table}[h] 
\centering
\makebox[\textwidth]{
\footnotesize
\sisetup{input-symbols=(),round-mode=places,round-precision=1}
\begin{tabular}{@{\extracolsep{5pt}} ll SSS} 
\toprule
\textbf{Predicted variable} & \textbf{Model} & \multicolumn{3}{c}{\textbf{Normalized RMSE}}\\ 
\midrule
&& \multicolumn{3}{c}{\textbf{Monthly resolution}} \\
\cmidrule(lr){3-5}
& & \multicolumn{1}{c}{$\bm{h=1}$} & \multicolumn{1}{c}{$\bm{h=12}$} & \multicolumn{1}{c}{$\bm{h=24}$}\\
\midrule 
  DAX 
   & Sentiment & 16.6 & 31.3 & 53.4 \\ 
   & Machine learning & 16.0 & 35.7 & 47.8 \\ 
   & Incl. dimensionality reduction & 16.0 & 34.7 & 51.9 \\[0.3em] 
  CDAX 
   & Sentiment & 17.1 & 31.4 & 54.5 \\ 
   & Machine learning & 16.6 & 36.2 & 48.9 \\ 
   & Incl. dimensionality reduction & 16.5 & 37.5 & 52.8 \\[0.3em] 
  STOXX 
   & Sentiment & 20.3 & 20.6 & 29.3 \\ 
   & Machine learning & 19.7 & 22.7 & 29.1 \\ 
   & Incl. dimensionality reduction & 20.2 & 23.5 & 29.2 \\ 
\midrule
&& \multicolumn{3}{c}{\textbf{Weekly resolution}} \\
\cmidrule(lr){3-5}
& & \multicolumn{1}{c}{$\bm{h=1}$} & \multicolumn{1}{c}{$\bm{h=52}$} & \multicolumn{1}{c}{$\bm{h=104}$}\\ 
\midrule 
  DAX 
   & Sentiment & 12.2 & 29.9 & 49.5 \\ 
   & Machine learning & 12.2 & 32.2 & 48.1 \\ 
   & Incl. dimensionality reduction & 12.1 & 33.3 & 45.6 \\[0.3em] 
  CDAX 
   & Sentiment & 12.0 & 29.7 & 51.5 \\ 
   & Machine learning & 12.0 & 33.1 & 51.1 \\ 
   & Incl. dimensionality reduction & 12.0 & 34.0 & 48.3 \\[0.3em] 
  STOXX 
   & Sentiment & 9.8 & 19.3 & 27.9 \\ 
   & Machine learning & 9.8 & 20.3 & 27.1 \\ 
   & Incl. dimensionality reduction & 9.8 & 21.8 & 25.4 \\
\bottomrule
\end{tabular}
}
\caption{Normalized RMSE for comparing the best-in-breed, text-based predictions.  
}
\label{tbl:nrmse}  
\end{table}

We now briefly validate the robustness of our results; that is, how sensitive the proposed machine learning approach is to the individual model choice. For this purpose, we report a detailed performance breakdown by model in the supplements. Whenever the machine learning approach in the summary table is strong in outperforming the lag-based baseline, we find similar outcomes when looking in the full palette of models. To quantify this effect, we computed the coefficient-of-variation across all estimated models. For instance, in the case of the monthly DAX, the coefficient-of-variation for the benchmarks computes to \num{0.183} in the nowcasting scenario, while it is lowered to \num{0.782} when using machine learning together with dimensionality reduction. Similar patterns arise in the two-year scenario where it drops from \num{0.032} to \num{0.022}.

\section{Discussion}
\label{sec:implications}

\subsection{Business implications for financial decision-making and decision-support}

The objective of this paper is to demonstrate the predictive power of approaches utilizing techniques from text mining in order to forecast stock indices. The predominant reason is that that trading recently witnessed a trends towards index ETFs. In this regard, the \emph{Financial Times} suggests that the relative trend towards ETFs as a form of passive investing continues to grow.\footnote{Financial Times. \emph{ETFs are eating the US stock market}. URL: \url{https://www.ft.com/content/6dabad28-e19c-11e6-9645-c9357a75844a}, last accessed April~29, 2018.} Our quantitative results thus aid practitioners, professional investors and managers.

Based on our findings, one can build algorithmic trading systems around our text-based prediction methodology, which are capable of executing potentially profitable trading strategies \cite{Gagnon.2013,Feuerriegel.2016}. In this regard, text mining in particular has recently received great traction and is propelling the automated interpretation of the linguistic content in corporate disclosures \cite{GrossKlumann.2011}. As an immediate implication, our research contributes to the stream of text-based trading and suggests the use of corporate disclosures in predicting stock indices, especially for long-term forecasts. This development is largely sustained by the growing volume and ease of access to unstructured narrative materials, thereby fundamentally advancing the methods of researchers in the field of financial forecasting.

A potential advantage of forecasting based on financial disclosures is the possibility of detecting market movements that are too complex for humans to identify. Yet our work also reveals several challenges. Among these is the veracity of financial news or, put differently, the information quality associated with verbal expressions. While financial news appears to be a significant driver of stock valuation \cite{Tetlock.2008}, there is still a large portion of unexplained variance. This noise component might be diminished by better forecasting techniques, although, a residual noise component might even be unpredictable, especially when signals are unclear or noisy \cite{Tetlock.2010}. 

From a mathematical point of view, our research setting is highly challenging because of the high-dimensional predictor matrix that can easily lead to overfitting. Here individual words are only weakly related the outcome variable and only their interplay accomplishes the desired prognostic power. As a remedy, our work lends to prescriptive guidelines for similar undertakings, as we compare different strategies for effective, text-based forecasting with wide data. This includes techniques for data-driven and knowledge-driven dimensionality reduction. We thereby contribute to the growing use of data mining techniques in financial forecasting \cite{Hu.2015}.

\subsection{Links to theory}

The semi-strong form of the efficient market hypothesis stipulates changes in financial valuation once novel information enters the market \cite{Fama.1965}. This is the case when corporations disclose ad~hoc announcements that subsequently drive a market response and thus trigger a direct change in the price variable. The corresponding forecasting performance draws purely upon novel information entering the market that consequently causes an adjustment of stock prices. Hence, the profits are not the result of arbitrage \cite{McLean.2016} and unlikely to diminish to zero-excess returns in the future.

Efficient markets \emph{\textquote{rule out the possibility of trading systems based only on (the assumed information set) that have expected profits or returns in excess of equilibrium profits or returns}} \cite[p.\,385]{Fama.1970}. However, research has also found empirical evidence supporting the predictability of stock prices \cite{Granger.1992,Hu.2015,Oztekin.2016,Timmermann.2004}. A recent meta-study lists 97 variables for which previous research has found a prognostic capacity of cross-sectional stock returns \cite{McLean.2016}. These variables include, for instance, analyst recommendations, turnover volume, bid-ask spread, investment decisions and tax levels. Our research yields findings analogous to prior work \emph{at stock level} that corporate disclosures have long-term prognostic capabilities \emph{at index level}. In this context, our work identifies highly complex and non-linear relationships between word choice and the future outlook of the economy.

\subsection{Limitations and potential for future research}

We provide evidence that financial disclosures \emph{per se} are linked to the economic outlook, but several aspects are left as potential avenues for future research. While this work demonstrates the prognostic capability of ad~hoc announcements in English and German for a range of stock markets, the task remains to repeat our analysis in other markets, such as investigating the interplay between Form 8-K filings in the \US and domestic stock indices. In our research design, the choice of regulatory disclosures entails several inherent advantages for financial forecasting, including their objectiveness, relevance to the market, concise format and short publication times. Nevertheless, one could also consider alternative text sources besides regulatory disclosures: for instance, social media, Internet stock message boards, newspaper releases or a combination thereof. Similarly, the analysis could be extended by including affective dimensions beyond sentiment or by analyzing the topic-specific reception. In addition, further effort is needed in order to obtain fully generative models and perform a rigorous model selection (for instance, see the test procedure in \cite{Giacomini.2006}).



\section{Conclusions} 
\label{sec:conclusion}

In this paper, we draw upon the efficient market hypothesis, which dictates that share valuations adjust to new information entering the market. We join theory and text mining based on which we test the ability of language, published in regulatory disclosures, to improve both short- and long-term forecasts of stock market indices. Our experiments reveal that text-based models perform at comparable levels relative to the baseline predictions for short-term forecasts and can outperform these baselines in terms of particularly challenging long-term forecasts.

To test the forecasting potential of our language-based data source, we utilize \num{20} years' worth of corporate disclosures mandated by German regulations. The individual disclosures are aggregated and processed into high-dimensional predictor matrices referring to the individual term frequencies. We then apply machine learning models, suited for high-dimensional problems, to forecast major German and European stock indices over multiple forecast horizons, up to 24 months ahead. We evaluate the forecasting errors of our text-based models against various benchmarks, including linear autogression and random forests, using lagged data as predictors.  With regard to the long-term forecasts experiments, the text-based models are able predict with lower forecast errors than the baseline models.




\bibliographystyle{model1-num-names}
\bibliography{literature}







\end{document}